\documentclass[conference]{IEEEtran}
\IEEEoverridecommandlockouts
\usepackage{cite}
\usepackage{amsmath,amssymb,amsfonts}
\usepackage{algorithmic}
\usepackage{graphicx}
\usepackage{textcomp}
\usepackage{xcolor}
\def\BibTeX{{\rm B\kern-.05em{\sc i\kern-.025em b}\kern-.08em
    T\kern-.1667em\lower.7ex\hbox{E}\kern-.125emX}}
\begin{document}

\title{Color Space-based HoVer-Net for Nuclei Instance Segmentation and Classification\\
}

\author{\IEEEauthorblockN{Hussam Azzuni}
\IEEEauthorblockA{\textit{Computer Vision Department} \\
\textit{Mohamed Bin Zayed University of Artificial Intelligence}\\
Abu Dhabi, United Arab Emirates \\
Hussam.Azzuni@mbzuai.ac.ae}
\and
\IEEEauthorblockN{Muhammad Ridzuan}
\IEEEauthorblockA{\textit{Machine Learning Department} \\
\textit{Mohamed Bin Zayed University of Artificial Intelligence}\\
Abu Dhabi, United Arab Emirates \\
Muhammad.Ridzuan@mbzuai.ac.ae}
\and
\IEEEauthorblockN{Min Xu}
\IEEEauthorblockA{\textit{Computer Vision Department} \\
\textit{Mohamed Bin Zayed University of Artificial Intelligence}\\
Abu Dhabi, United Arab Emirates \\
Min.Xu@mbzuai.ac.ae}
\and
\IEEEauthorblockN{Mohammad Yaqub}
\IEEEauthorblockA{\textit{Computer Vision Department} \\
\textit{Mohamed Bin Zayed University of Artificial Intelligence}\\
Abu Dhabi, United Arab Emirates \\
Mohammad.Yaqub@mbzuai.ac.ae}
}

\maketitle

\begin{abstract}
Nuclei segmentation and classification is the first and most crucial step that is utilized for many different microscopy medical analysis applications. However, it suffers from many issues such as the segmentation of small objects, imbalance, and fine-grained differences between types of nuclei. In this paper, multiple different contributions were done tackling these problems present. Firstly, the recently released "ConvNeXt" was used as the encoder for HoVer-Net model since it leverages the key components of transformers that make them perform well. Secondly, to enhance the visual differences between nuclei, a multi-channel color space-based approach is used to aid the model in extracting distinguishing features. Thirdly, Unified Focal loss (UFL) was used to tackle the background-foreground imbalance. Finally, Sharpness-Aware Minimization (SAM) was used to ensure generalizability of the model. Overall, we were able to outperform the current state-of-the-art (SOTA), HoVer-Net, on the preliminary test set of the CoNiC Challenge 2022 by 12.489\% mPQ+.

\end{abstract}


\section{Introduction}

Nuclei segmentation and classification is an important step used in many different applications such as cell type classification, cell counting, and phenotype analysis \cite{ref_article1}. However, it suffers from multiple challenges such as the presence of clustered nuclei, the fine differences between different types of nuclei, and finally the imbalance between different classes and between the background and foreground classes.

Convolution Neural Networks (CNNs) have mostly been used for nuclei segmentation and classification. U-Net \cite{b1} is the de-facto method when it comes to object segmentation. BiO-Net \cite{b2} improved on U-Net by reusing the available blocks in U-Net recurrently by introducing bi-directional skip connections between the encoder and the decoder. This allows improving the performance without increasing the number of parameters. Multiple CNN methods have been used to tackle one of the biggest issues in nuclei segmentation which is the presence of clustered nuclei such as BRP-Net \cite{b3}, which consists of two stages: one, to obtain nuclei proposals and two, to use these proposals for prediction. Finally, HoVer-Net \cite{b4} uses the distance between the horizontal and vertical nuclear pixels and its centre of mass as important information to separate those clustered nuclei.

Recently, "Lizard" \cite {b10}, a large-scale nuclear instance segmentation and classification dataset, has been released which consists of almost half-a-million nuclei across 4,981 images. It includes six different classes which are neutrophils, epithelial, lymphocyte, plasma, eosinophil, and connective nuclei. However, the number of instances and their shape morphology can differ. Our contributions are as follows:

\begin{enumerate}
    \item Utilizing ConvNeXt, which is a modernized version of CNNs that uses the key components that made transformers work well.
    \item Usage of multi-channel color space-based input that uses different color spaces to provide important information for the model.
    \item Improving the generalization of the model using different augmentation techniques, loss function, and Sharpness-Aware Minimization.
\end{enumerate}

\section{Data preprocessing and augmentation}

\subsection{Contrast enhancement}

Contrast Limited Adaptive Histogram Equalization (CLAHE) \cite{b5} is used to improve contrast of the input images, making it easier for the model to extract meaningful information. Since CLAHE is applied to single channels, the RGB image was split into its three channels: R, G, and B. Then, CLAHE was applied to every channel individually, and the channels are merged again before being fed into the network.

\begin{table*}[t]
\caption{Preliminary test results based on CoNiC leaderboard. All the experiments are done using modified ConvNeXt-small as the encoder}
\begin{center}
\begin{tabular}{|c|c|c|c|c|c|c|c|c|c|}
\hline
\textbf{Experiment} & \textbf{mPQ+} &  \textbf{PQ} &  \textbf{PQ+} &  \textbf{PQ+ - pla} &  \textbf{PQ+ - neu} &  \textbf{PQ+ - epi} &  \textbf{PQ+ - lym} &  \textbf{PQ+ - eos} &  \textbf{PQ+ - con} \\
\hline
HoVer-Net & 0.29558 & 0.55031 & 0.54385 & 0.28953 & 0.00904 & 0.50457 & 0.41308 & 0.21342 & 0.34384 \\
\hline
ConvNeXt-small & 0.35216 & 0.56323 & 0.56228 & 0.24046 & 0.11744 & 0.54114 & 0.44712 & 0.40283 & 0.36397\\
\hline
+ Affine & 0.40360 & \bf 0.58543 & \bf 0.58322 & 0.40498 & 0.12612 & \bf 0.56994 & 0.44350 & 0.44759 & \bf 0.42950\\
\hline
+ 6-channel + SAM + 4UFL & \bf 0.42047 & 0.57244 & 0.57113 & \bf 0.44752 & \bf 0.14807 & 0.56081 & \bf 0.44832 & \bf 0.49797 & 0.42012\\
\hline
\end{tabular}
\label{tab1}
\end{center}
\end{table*}

\subsection{Multi-channel color-stacking}
\begin{figure}[t]
\centerline{\includegraphics[scale = 0.22]{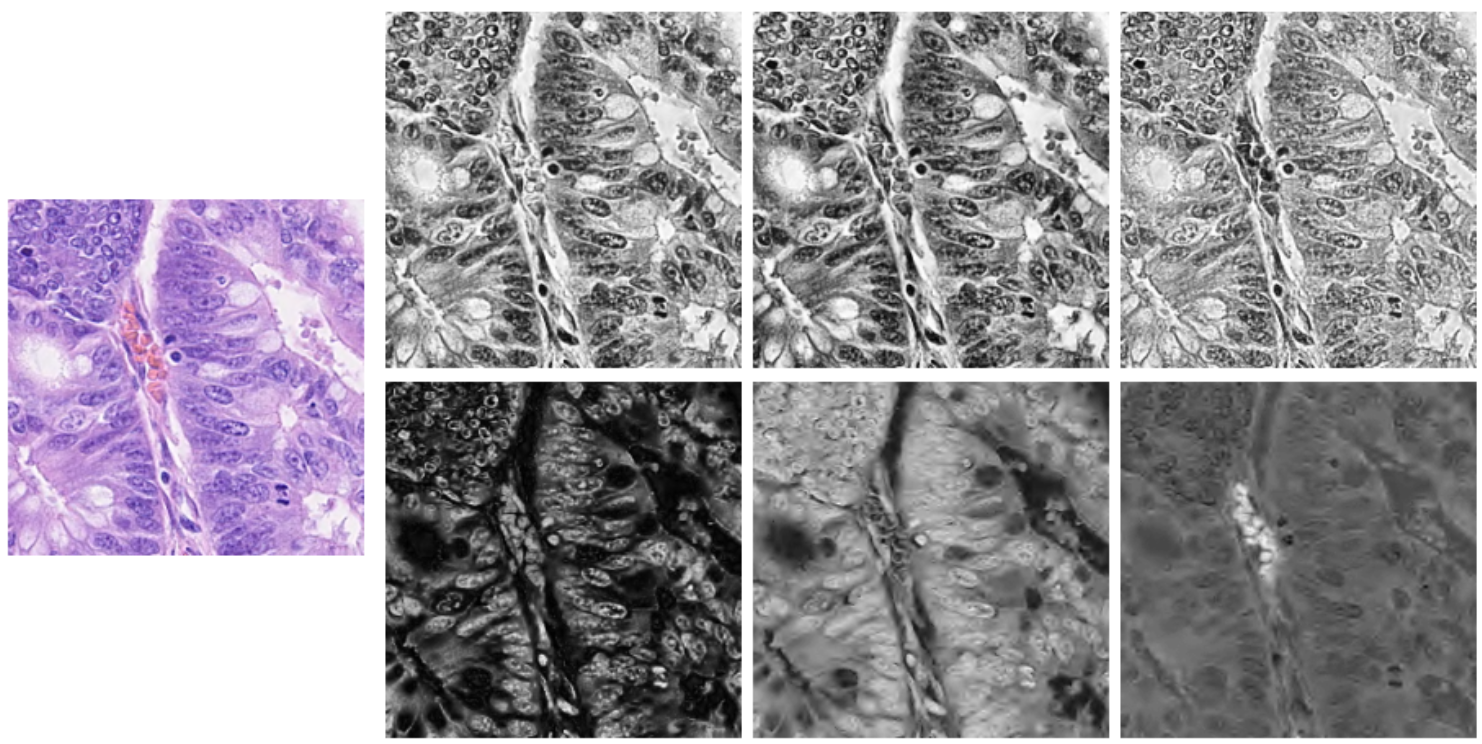}}
\caption{Example input image with six multi-channel inputs. From top left to bottom right: channels B, G and R from RGB, S from HSV, and Cr and Cb from YCrCb.}
\label{fig}
\end{figure}

Multi-channel color-stacking that combines the different channels from various color spaces is effective in detecting the fine features that help classify the different types of nuclei. We found that saturation (channel 2) in HSV color space, red-yellow and blue-yellow chrominances (channels 2 and 3) in YCrCb color space contained distinctive visual information not available in the RGB color space. Thus, these channels were merged to create a 6-channel input shown in Fig. \ref{fig}.

\subsection{Augmentation}
Various augmentations were used to diversify the histopathology input images for the model to generalize better. Based on the implementation of HoVer-Net, varying the images by inputting noise (Gaussian blur, median blur, or additive Gaussian noise), and color space values were used for better generalization. Additionally, we also applied $\pm$5 degree shear, $\pm$20\% scaling to account for the different morphological variations of the different types of nuclei \cite{b4} .

\section{Method}

HoVer-Net \cite{b4} was taken as our baseline architecture for instance segmentation and classification. This model consists of a Preact-ResNet50 encoder and a 3-branch decoder. In our implementation, we replaced the encoder with the recently released "ConvNeXt". Our best performing model was trained using AdamW optimizer and a stepLR scheduler with a step size of 25.

\subsection{ConvNeXt}
ConvNeXt \cite{b8} aims to improve on classical CNNs by utilizing transformer design components to further refine CNNs to reach its maximum potential. ConvNeXt-small was modified by replacing the 'patchify' layer with a stride of 4, into a 7$\times$7 convolution with a stride of 1 to avoid downsampling the image since the model is dealing with small objects.

\subsection{Loss function} 
Unified Focal loss (UFL) \cite{b7} is based on generalizing both Dice and cross-entropy loss to tackle class imbalance. Given the abundant foreground-background class imbalance, both UFL and Dice loss were used on the tp and np branches of the HoVer-Net decoder to optimize the model, leading to a better convergence. 

\subsection{Sharpness-Aware Minimization}
Sharpness-Aware Minimization (SAM) \cite{b9} is an optimization technique that ensures the neighbourhood parameters have similar low losses, which eventually results in having a more generalized model. This is done by formulating the optimization as a min-max problem making the optimizer perform effectively.

\section{Results}
Table \ref{tab1} shows the results based on the preliminary test set of CoNiC Challenge 2022 \cite{b11}. Firstly, changing the original encoder with the modified ConvNeXt-small improved the performance in most classes, especially eosinophils, resulting in a 5.658\% increase in mPQ+. Later on, the addition of multiple affine transformation such as shearing with $\pm$ 5\%, and scaling with a $\pm$20\% gave a major improvement on most classes, resulting in a 5.144\% increase on top of the previous change. Finally, the usage of 6-channel input, SAM, and weighted Unified Focal loss improved the model even further by 1.687\%, resulting in an overall increase of 12.489\% mPQ+ from the initial baseline.

\section{Conclusion}
In this work, we proposed the usage of ConvNeXt as a replacement of the current Preact-ResNet50 encoder since it provides an overall better performance in nuclei segmentation and classification. Furthermore, the usage of affine transformations, which provides the model with much more variety in nuclei shapes and sizes, together with SAM, lead to a more generalizable model. Thirdly, the usage of multiple channels from different color spaces provided the model with extra effective information that improved the performance in different nuclei classes. Finally, the usage of both unified focal loss and dice loss is crucial given the imbalanced nature between the foreground and the background.


\end{document}